\documentclass[pra,showkeys,twocolumn]{revtex4}%
\pdfoutput=1
\usepackage{amsfonts}
\usepackage{amsmath}
\usepackage{amssymb}
\usepackage{graphicx}
\usepackage{verbatim}%
\providecommand{\U}[1]{\protect\rule{.1in}{.1in}}
\newtheorem{theorem}{Theorem}

\newtheorem{definition}[theorem]{Definition}

\begin{document}
\preprint{ }
\title{Coherent Communication with Linear Optics}
\author{Mark M. Wilde}
\author{Todd A. Brun}
\affiliation{Communication Sciences Institute, Department of Electrical Engineering,
University of Southern California, Los Angeles, California 90089}
\author{Jonathan P. Dowling}
\author{Hwang Lee}
\affiliation{Hearne Institute for Theoretical Physics, Department of Physics and Astronomy,
Louisiana State University, Baton Rouge, Louisiana 70803}

\begin{abstract}
We show how to implement several continuous-variable coherent protocols with
linear optics. Noise can accumulate when implementing each coherent protocol
with realistic optical devices. Our analysis bounds the level of noise
accumulation. We highlight the connection between a coherent channel and a
nonlocal quantum nondemolition interaction and give two new protocols that
implement a coherent channel. One protocol is superior to a previous method
for a nonlocal quantum nondemolition interaction because it requires fewer
communication resources. We then show how continuous-variable coherent
superdense coding implements two nonlocal quantum nondemolition interactions
with a quantum channel and bipartite entanglement. We finally show how to
implement continuous-variable coherent teleportation experimentally and
provide a way to verify the correctness of its operation.

\end{abstract}
\volumeyear{2007}
\volumenumber{ }
\issuenumber{ }
\eid{ }
\date{}
\received[Received text]{}

\revised[Revised text]{}

\accepted[Accepted text]{}

\published[Published text]{}

\startpage{1}
\endpage{ }
\maketitle

\section{Introduction}

Quantum information theory employs a plethora of resources that aid in
communication \cite{ieee1998bennett}. Classical communication and entanglement
partner together in quantum teleportation to create a noiseless quantum
channel between a sender and receiver \cite{PhysRevLett.70.1895}. Quantum
communication and entanglement partner in superdense coding to create two
noiseless classical channels between sender and receiver
\cite{PhysRevLett.69.2881}. It is possible to combine the resources of
classical communication, quantum communication, and entanglement in a variety
of ways to send classical or quantum information \cite{prl2004dev,arx2005dev}.

Harrow introduced a new element to quantum information theory: the coherent
bit channel \cite{prl2004harrow}. The coherent bit channel is the following
isometry%
\begin{equation}
\left\vert x\right\rangle ^{A}\rightarrow\left\vert x\right\rangle
^{A}\left\vert x\right\rangle ^{B}:x\in\left\{  0,1\right\}  .
\end{equation}
An arbitrary qubit $\left\vert \psi\right\rangle ^{A}=\alpha\left\vert
0\right\rangle ^{A}+\beta\left\vert 1\right\rangle ^{A}$ becomes the following
state%
\begin{equation}
\alpha\left\vert 0\right\rangle ^{A}\left\vert 0\right\rangle ^{B}%
+\beta\left\vert 1\right\rangle ^{A}\left\vert 1\right\rangle ^{B}%
\end{equation}
after sending it through the cobit channel. The operation of the cobit channel
is similar to the \textquotedblleft quantum encoder\textquotedblright%
\ \cite{PhysRevA.64.062311}.

The coherent bit channel or \textquotedblleft cobit\textquotedblright\ channel
is fundamentally different from classical communication, quantum
communication, or entanglement, but has connections to all three of the above
resources. Harrow originally interpreted the cobit channel as a coherent
version of a classical channel. He defined it as a quantum feedback operation
in which the sender essentially becomes the environment in a dephasing
channel. The cobit channel is similar to quantum communication because a
quantum state encodes the transmitted message. The cobit channel thus
maintains coherent superpositions of quantum states. This interpretation is
the root of the cobit channel's name \cite{qic2005harrow}. The cobit channel
has links to entanglement because protocols that employ it typically consume
less entanglement than their incoherent counterparts \cite{prl2004harrow}.
Some protocols such as coherent teleportation even generate entanglement in a
certain sense \cite{prl2004harrow}.

The cobit channel proves useful in constructing coherent implementations of
established quantum communication protocols. It gives coherent versions of
remote state preparation, teleportation, dense coding, distributed unitaries
\cite{prl2004harrow}, and entanglement-assisted quantum codes
\cite{science2006brun,arx2006brun}. The cobit channel is also useful as an
intermediate step in several quantum information-theoretic proofs
\cite{arx2006devyard,arx2007yard} and gives the capacity of a unitary gate
\cite{arx2005harrow}. Both the direct construction of coherent protocols and
the employment of the cobit channel in quantum information-theoretic proofs
affirm its status as a fundamental primitive for quantum communication.

A theory of a continuous-variable coherent channel recently emerged
\cite{wilde:060303}. This theory incorporates the effect of finitely squeezed
states in continuous-variable quantum information processing
\cite{book2003braunstein,revmod2005braunstein}. The continuous-variable
coherent channel produces coherent versions of continuous-variable
teleportation \cite{prl1998braunstein}\ and\ continuous-variable superdense
coding \cite{optb1999ban,pra2000braunstein}. The continuous-variable coherent
channel should prove useful as a resource for continuous-variable
communication protocols or in proving various capacities for a
continuous-variable quantum channel.

Theoretical quantum information needs experiment to validate its predictions.
The field of quantum information will be successful only through incremental
demonstrations of quantum communication protocols. Furusawa, et al., pushed
the field of quantum information ahead by performing a continuous-variable
teleportation experiment \cite{science1998furusawa}\ that validated the
predictions in Ref.~\cite{prl1998braunstein}. The advantage of the experiment
is that its implementation requires linear optics---passive optical elements,
offline squeezers, homodyne detection, feedforward classical signaling, and
conditional displacements. Several experimentalists have since implemented
continuous-variable superdense coding using linear optics
\cite{mizuno:012304,PhysRevLett.88.047904}. The major advantage that
continuous-variable quantum information possesses right now is the ease with
which experimentalists can create and control the modes of the electromagnetic
field. Continuous-variable quantum information therefore provides an ideal
testbed for validating theoretical predictions.

In this work, we give an explicit proposal for a linear-optical experiment to
implement the coherent protocols previously outlined \cite{wilde:060303}. Our
proposals make extensive use of the scheme by Filip, Marek, and Andersen
(FMA)\ for a quantum nondemolition interaction \cite{filip:042308}. We give a
noise model that bounds the performance of the protocols when using FMA's
scheme. The schemes and analysis in this work should provide a clear path for
implementing the coherent channel.

We also provide an additional interpretation of a coherent channel as a
nonlocal quantum nondemolition interaction. This interpretation suggests that
the coherent channel is useful for distributed quantum computation.

We give two new protocols that implement a continuous-variable coherent
channel. One of these protocols (CCAECC) requires fewer communication
resources than a protocol previously outlined by Filip \cite{filip:052313}.

We structure our work as follows. Our review in Section~\ref{sec:def-review}
includes the definition of the discrete-variable and continuous-variable
coherent channels. Our review in Section~\ref{sec:FMA-review} gives FMA's
scheme for a quantum nondemolition\ interaction \cite{filip:042308} because
our linear-optical schemes employ this technique. In
Section~\ref{sec:coherent-imps}, we outline three different ways to implement
a continuous-variable coherent channel with linear optics. Coherent superdense
coding is the last of these implementations, it is the most efficient in its
use of resources, and it is equivalent to implementing two nonlocal quantum
nondemolition interactions. Section~\ref{sec:coh-tele}\ shows how to implement
continuous-variable coherent teleportation with linear optics. We finally
provide a loss analysis for each coherent protocol. These losses are due to
finite squeezing, inefficient photodetectors, and inefficient feedforward control.

\section{Definitions}

\label{sec:def-review}We first review the discrete-variable coherent channel.
Harrow defines a classical bit channel from a sender Alice to a receiver Bob
as the following isometry:%
\begin{equation}
\left\vert x\right\rangle ^{A}\rightarrow\left\vert x\right\rangle
^{E}\left\vert x\right\rangle ^{B}:x\in\left\{  0,1\right\}  .
\end{equation}
The channel is classical because the environment correlates with Alice's
state. Thus the channel does not maintain coherent superpositions. The
coherent channel supposes that Alice regains the environment's state---earning
its alias as the quantum feedback operation. It is the following isometry:%
\begin{equation}
\left\vert x\right\rangle ^{A}\rightarrow\left\vert x\right\rangle
^{A}\left\vert x\right\rangle ^{B}:x\in\left\{  0,1\right\}  .
\end{equation}
The cobit channel is similar to classical copying because it copies the basis
states while maintaining coherent superpositions.

The above definition tempts one to define a continuous-variable coherent
channel as the following map:%
\begin{equation}
\left\vert x\right\rangle ^{A}\rightarrow\left\vert x\right\rangle
^{A}\left\vert x\right\rangle ^{B}:x\in\mathbb{R}.
\end{equation}
The above states are position-quadrature or momentum-quadrature eigenstates.
The problem with the above definition is that it requires an infinite amount
of energy to implement. It requires infinite energy to copy the eigenstates
because they have continuous degrees of freedom. The above coherent channel is
therefore an idealized limit. Consider the effect of sending an arbitrary
state $\left\vert \psi\right\rangle =\int\psi\left(  x\right)  \ \left\vert
x\right\rangle \ dx$ through the ideal coherent channel. The resulting state
$\int\psi\left(  x\right)  \ \left\vert x\right\rangle ^{A}\left\vert
x\right\rangle ^{B}\ dx$ is not normalizable and thus has infinite energy. We
therefore set aside the ideal coherent coherent channel, and instead use a
definition that allows for finitely-squeezed states.

We turn to the Heisenberg picture to formulate the definition in terms of the
quadrature operators of the electromagnetic field. The definition has a
parameter $\epsilon$ that determines the performance of the coherent channel
and indicates how much squeezing is present in the channel.

\begin{definition}
\label{def:coherent-channel}An $\epsilon$-approximate position-quadrature
coherent channel $\tilde{\Delta}_{X}$\ is any mechanism by which one mode
transforms to two modes:%
\begin{equation}%
\begin{bmatrix}
\hat{x}_{A} & \hat{p}_{A}%
\end{bmatrix}
\ \underrightarrow{\tilde{\Delta}_{X}}\
\begin{bmatrix}
\hat{x}_{A^{\prime}} & \hat{p}_{A^{\prime}} & \hat{x}_{B^{\prime}} & \hat
{p}_{B^{\prime}}%
\end{bmatrix}
.
\end{equation}
It maintains the canonical commutation relations:%
\begin{equation}
\left[  \hat{x}_{A^{\prime}},\hat{p}_{A^{\prime}}\right]  =\left[  \hat
{x}_{B^{\prime}},\hat{p}_{B^{\prime}}\right]  =i.
\end{equation}
It maps the input quadrature operators to the output quadrature operators as
follows:%
\begin{align}
\hat{x}_{A^{\prime}}  &  =\hat{x}_{A},\label{eq:position-retaining}\\
\hat{x}_{B^{\prime}}  &  =\hat{x}_{A}+\hat{x}_{\Delta_{X}}%
,\label{eq:position-copying}\\
\hat{p}_{A^{\prime}}  &  =\hat{p}_{A}+\hat{p}_{\Delta_{X}},
\label{eq:back-action}%
\end{align}
where%
\begin{equation}
\left\langle \hat{x}_{\Delta_{X}}\right\rangle =\left\langle \hat{p}%
_{\Delta_{X}}+\hat{p}_{B^{\prime}}\right\rangle =0.
\end{equation}
The condition in (\ref{eq:position-copying}) indicates that the position
quadrature $\hat{x}_{A}$ copies to mode $B^{\prime}$ with the addition of some
noise $\hat{x}_{\Delta_{X}}$. The condition in (\ref{eq:back-action})
indicates back action in the momentum quadrature $\hat{p}_{A^{\prime}}$. The
parameter $\epsilon$ bounds the performance of the channel by bounding the
noise terms as follows:%
\begin{equation}
\langle\hat{x}_{\Delta_{X}}^{2}\rangle\leq\epsilon,\ \ \langle\left(  \hat
{p}_{\Delta_{X}}+\hat{p}_{B^{\prime}}\right)  ^{2}\rangle\leq\epsilon.
\label{eq:coherent-performance}%
\end{equation}

\end{definition}

A coherent channel is similar to a nonlocal quantum nondemolition interaction.
Suppose that Alice has a mode $A$, Bob has a mode $B$, and that Alice and Bob
are spacelike separated. A position-quadrature nonlocal quantum nondemolition
interaction implements the following transformation%
\begin{align}
\hat{x}_{A}  &  \rightarrow\hat{x}_{A},\ \ \hat{p}_{A}\rightarrow\hat{p}%
_{A}-g\hat{p}_{B},\label{eq:qnd-def}\\
\hat{x}_{B}  &  \rightarrow\hat{x}_{B}+g\hat{x}_{A},\ \ \hat{p}_{B}%
\rightarrow\hat{p}_{B},\nonumber
\end{align}
where $g$ is the gain of the interaction. A coherent channel has more
restrictions than a nonlocal quantum nondemolition interaction. The coherent
channel begins with only one mode. It requires the final gain of the
interaction to be unity. It further requires both output position quadratures
to be $\epsilon$-close in mean-squared distance. A quantum nondemolition
interaction merely requires the second quadrature to have information about
the first quadrature. However, it is still useful to consider this connection.
Observe the output quadratures in Definition~\ref{def:coherent-channel}. The
original position quadrature $\hat{x}_{A}$\ copies to the output quadrature
$\hat{x}_{B^{\prime}}$. The output momentum-quadrature $\hat{p}_{A^{\prime}}$
also has back-action noise encapsulated in the operator $\hat{p}_{\Delta_{X}}%
$. One party possesses the first mode and the other party possesses the second
mode. This behavior is similar to the behavior in a nonlocal quantum
nondemolition interaction.

The definition for a momentum-quadrature coherent channel is similar to
Definition~\ref{def:coherent-channel}. It is useful to have a
momentum-quadrature coherent channel definition to observe the similarities
between coherent protocols \cite{wilde:060303} and their incoherent
counterparts \cite{prl1998braunstein,optb1999ban,pra2000braunstein}.

\begin{definition}
An $\epsilon$\textit{-approximate momentum-quadrature coherent channel}
$\tilde{\Delta}_{P}$ performs the following transformation with conditions:%
\begin{align}%
\begin{bmatrix}
\hat{x}_{A} & \hat{p}_{A}%
\end{bmatrix}
\  &  \underrightarrow{\tilde{\Delta}_{P}}\
\begin{bmatrix}
\hat{x}_{A^{\prime\prime}} & \hat{p}_{A^{\prime\prime}} & \hat{x}%
_{B^{\prime\prime}} & \hat{p}_{B^{\prime\prime}}%
\end{bmatrix}
,\label{eqn:cond-m-conat-1}\\
\left[  \hat{x}_{A^{\prime\prime}},\hat{p}_{A^{\prime\prime}}\right]   &
=\left[  \hat{x}_{B^{\prime\prime}},\hat{p}_{B^{\prime\prime}}\right]
=i,\nonumber\\
\hat{p}_{A^{\prime\prime}}  &  =\hat{p}_{A},\nonumber\\
\hat{p}_{B^{\prime\prime}}  &  =\hat{p}_{A}+\hat{p}_{\Delta_{P}},\nonumber\\
\hat{x}_{A^{\prime\prime}}  &  =\hat{x}_{A}+\hat{x}_{\Delta_{P}},\nonumber\\
\left\langle \hat{p}_{\Delta_{P}}\right\rangle  &  =\left\langle \hat
{x}_{\Delta_{P}}+\hat{x}_{B^{\prime\prime}}\right\rangle =0,\nonumber\\
\langle\hat{p}_{\Delta_{P}}^{2}\rangle,  &  \ \ \langle\left(  \hat{x}%
_{\Delta_{P}}+\hat{x}_{B^{\prime\prime}}\right)  ^{2}\rangle\leq
\epsilon.\nonumber
\end{align}

\end{definition}

A coherent channel is useful for implementing a coherent teleportation
protocol. The parameter $\epsilon$ indicates the performance of the coherent
channel in a coherent teleportation protocol \cite{wilde:060303}.
Continuous-variable teleportation schemes have several different performance
bounds \cite{prl1998braunstein,PhysRevA.64.010301,caves:040506}. The
performance bounds are relevant lower bounds for the average fidelity of
teleporting an arbitrary coherent state. The preparation-and-measurement limit
is 1/2 \cite{prl1998braunstein} and the no-cloning limit is 2/3
\cite{PhysRevA.64.010301,caves:040506}. The coherent-state-averaged fidelity
for coherent teleportation exceeds 1/2 if $\epsilon<1$. The original paper on
continuous-variable coherent communication only considered this bound of 1/2
\cite{wilde:060303}. Examination of the equations in \cite{wilde:060303}
indicates that the coherent-state-averaged fidelity exceeds 2/3 if
$\epsilon<1/2$.

\section{FMA's Quantum Nondemolition Interaction}

\label{sec:FMA-review}The coherent protocols originally outlined in
Ref.~\cite{wilde:060303} require a quantum nondemolition interaction for their
implementation. Quantum nondemolition interactions typically involve an online
nonlinear interaction such as a Kerr medium \cite{nat1998grang}. These online
nonlinear interactions are difficult to control experimentally.%

\begin{figure}
[pt]
\begin{center}
\includegraphics[
natheight=4.152800in,
natwidth=5.124900in,
height=1.4313in,
width=1.7642in
]%
{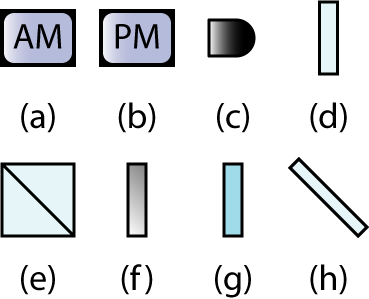}%
\caption{(Color online)\ We outline the elements used in our optical circuits.
(a) An amplitude modulator displaces the position quadrature of an optical
mode. A classical control signal controls the amount of displacement. (b) A
phase modulator kicks the momentum quadrature of an optical mode. A classical
signal also controls it. (c) A photodetector. (d) A phase shifter. (e) A
polarizing beam splitter. (f) A mirror. (g) A half-wave plate with variable
transmittivity. (h) A beam splitter.}%
\label{fig:legend}%
\end{center}
\end{figure}
FMA provided an experimental proposal for implementing both a squeezer and a
quantum nondemolition interaction with linear optics \cite{filip:042308}.
Experimentalists have implemented both the squeezing transformation
\cite{arx2007furusawa} and the quantum nondemolition\ interaction
\cite{conf2007furusawa} with reasonable performance.

\begin{figure}
[ptb]
\begin{center}
\includegraphics[
natheight=14.400000in,
natwidth=17.947399in,
height=2.4163in,
width=3.0096in
]%
{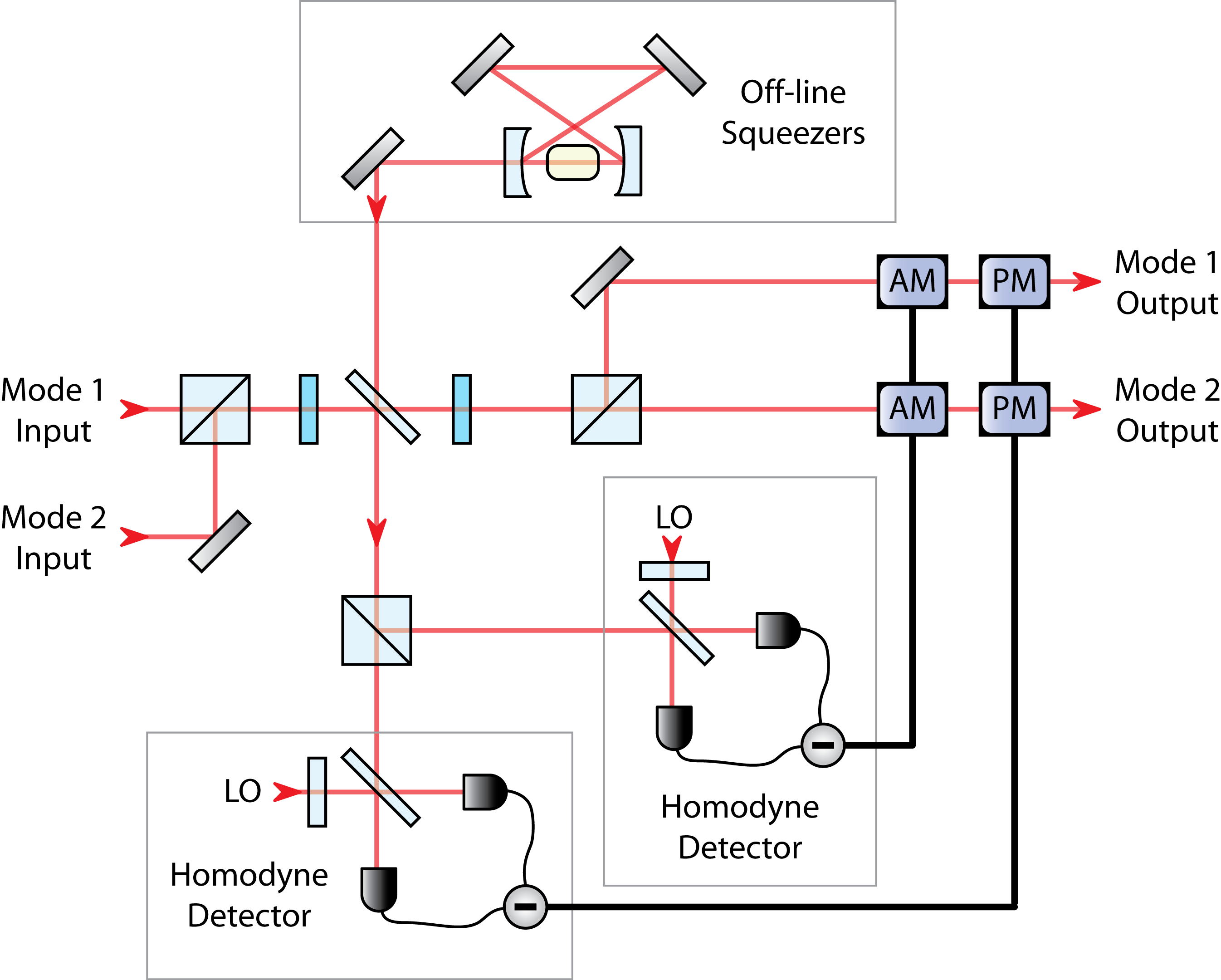}%
\caption{(Color online) FMA's linear-optical scheme for a quantum
nondemolition interaction \cite{filip:042308}. LO is an abbreviation for
\textquotedblleft local oscillator.\textquotedblright\ The two input modes are
orthogonally polarized. The circuit uses two offline squeezers, homodyne
detection, and feedforward control to give a quantum nondemolition interaction
with unity transfer gain.}%
\label{fig:filip-QND}%
\end{center}
\end{figure}

Figure~\ref{fig:legend} highlights the elements used in a linear-optical
circuit. Figure~\ref{fig:filip-QND}\ gives the optical circuit for
implementing FMA's measurement-induced quantum nondemolition interaction.
FMA's scheme uses two offline squeezed modes, three homodyne measurements, and
feedforward control.

FMA's scheme is similar in spirit to the Knill, Laflamme, and Milburn scheme
for discrete variables \cite{nat2001klm} though FMA's scheme has the advantage
that it is deterministic rather than probabilistic.

The scheme for a quantum nondemolition interaction is valuable for any
continuous-variable quantum computation or communication device. A recent work
used it in an algorithm for constructing linear-optical encoding circuits for
continuous-variable quantum error correction \cite{arx2007wilde}. We use it in
all of our coherent protocols below.%

We review the operation of FMA's scheme. The circuit uses two offline
squeezers to implement a quantum nondemolition interaction with unity transfer
gain. It begins with a polarizing beam splitter combining the two inputs into
one spatial mode. A half-wave plate with transmittivity $T_{1}$\ mixes the two
polarization modes. A beamsplitter then combines the spatial mode and the
outputs of two offline squeezers. The outputs of the offline squeezers are two
orthogonally polarized modes squeezed in conjugate quadratures $\hat{x}_{B}$
and $\hat{p}_{A}$. We measure the outputs of the beamsplitter with two
homodyne detectors. The two phase shifters for the local oscillators control
which quadrature we measure and thus which quadrature is the nondemolition
variable. We send the rightward output of the beamsplitter through a half-wave
plate with transmittivity $T_{2}$. FMA require setting the parameters $T_{1}$
and $T_{2}$ as follows%
\begin{equation}
T_{1}=1/\left(  1+T\right)  ,\ \ \ \ T_{2}=T/\left(  1+T\right)  .
\end{equation}
to have unity transfer gain. A polarizing beamsplitter separates the output of
the half-wave plate into two spatial modes. We finally perform modulation of
both modes using the results of the homodyne detection.

The above operations transform the input quadrature observables $\hat{x}_{1}$,
$\hat{p}_{1}$, $\hat{x}_{2}$, $\hat{p}_{2}$ to the final output quadrature
observables $\hat{x}_{1}^{\prime}$, $\hat{p}_{1}^{\prime}$, $\hat{x}%
_{2}^{\prime}$, $\hat{p}_{2}^{\prime}$ as follows:%
\begin{align}
\hat{x}_{1}^{\prime}  &  =\hat{x}_{1}-\sqrt{\alpha}\hat{x}_{0}-\sqrt{\beta
}\hat{x}_{B},\nonumber\\
\hat{p}_{1}^{\prime}  &  =\hat{p}_{1}-\left(  \frac{1}{\sqrt{T}}-\sqrt
{T}\right)  \hat{p}_{2}+\sqrt{\alpha/T}\hat{p}_{0}+\sqrt{T\beta}\hat{p}%
_{A},\nonumber\\
\hat{x}_{2}^{\prime}  &  =\hat{x}_{2}+\left(  \frac{1}{\sqrt{T}}-\sqrt
{T}\right)  \hat{x}_{1}-\sqrt{\alpha/T}\hat{x}_{0}+\sqrt{T\beta}\hat{x}%
_{B},\nonumber\\
\hat{p}_{2}^{\prime}  &  =\hat{p}_{2}-\sqrt{\alpha}\hat{p}_{0}+\sqrt{\beta
}\hat{p}_{A}.
\end{align}
Parameter $T$ controls the strength of the interaction. The parameters
$\alpha$ and $\beta$ determine the efficiency of the interaction. The
quadratures $\hat{x}_{0}$ and $\hat{p}_{0}$ are independent and commuting
vacuum contributions. We set $T=\left(  3-\sqrt{5}\right)  /2$ for our
purposes throughout this work. This setting ensures unity gain for every
quantum nondemolition interaction so that%
\begin{equation}
\frac{1}{\sqrt{T}}-\sqrt{T}=1.
\end{equation}

The above method adds noise to each mode in both quadrature observables. The
parameters $\alpha$ and $\beta$ are as follows%
\begin{align}
\beta &  =\frac{1-T}{1+T}=\frac{-1+\sqrt{5}}{5-\sqrt{5}},\\
\alpha &  =\frac{\left(  1-T\right)  \left(  1-\eta\right)  }{\left(
1+T\right)  \eta}=\beta\left(  \frac{1-\eta}{\eta}\right)  .
\end{align}
where $\eta$ is the efficiency of the photodetectors. Let $\eta_{F}$ denote
the total efficiency of FMA's scheme. We define $\eta_{F}$ to be a bound on
the second moment of the added noise for the quantum nondemolition interaction%
\begin{equation}
\eta_{F}=\beta\left(  \left(  1-\eta\right)  /\eta T+e^{-2r}\right)  ,
\label{eq:QND-noise-bound}%
\end{equation}
where $r$ is the strength of the offline squeezers. We use this bound
throughout our work to quantify the performance of FMA's quantum nondemolition interaction.

\section{Practical Implementations of a Coherent Channel}

\label{sec:coherent-imps}Definition~\ref{def:coherent-channel}\ for a
continuous-variable coherent channel is rather abstract. It gives a set of
conditions that a coherent channel must satisfy and is also broad in its
scope. The conditions are necessary to implement a coherent superdense coding
protocol and are sufficient to implement a coherent teleportation protocol
\cite{wilde:060303}. We illuminate the abstract definition of a coherent
channel in this section by providing several ways to implement it. We also
further highlight the connection between a coherent channel and a nonlocal
quantum nondemolition interaction.

\subsection{Quantum Nondemolition Interaction}

The simplest method of implementing a coherent channel is with a quantum
nondemolition interaction. Suppose that a sender Alice possesses a mode $A$
that she wants to send through a coherent channel. She creates a
position-squeezed ancilla mode $B$\ with squeezing strength $r$. The
Heisenberg-picture observables corresponding to the two-mode state are as
follows%
\begin{equation}
\hat{x}_{A},\ \ \hat{p}_{A},\ \ \hat{x}_{B}^{\left(  0\right)  }%
e^{-r},\ \ \hat{p}_{B}^{\left(  0\right)  }e^{r},
\end{equation}
where the observables $\hat{x}_{A}$ and $\hat{p}_{A}$ are her original modes
and the observables $\hat{x}_{B}^{\left(  0\right)  }$ and $\hat{p}%
_{B}^{\left(  0\right)  }$ have the fluctuations of the vacuum. It is implicit
throughout this work that any quadrature with $\left(  0\right)  $ in the
superscript has the fluctuations of the vacuum. Alice performs a quantum
nondemolition interaction on her two modes so that the observables evolve as
follows:%
\begin{equation}
\hat{x}_{A},\ \ \hat{p}_{A}-\hat{p}_{B}^{\left(  0\right)  }e^{r},\ \ \hat
{x}_{B}^{\left(  0\right)  }e^{-r}+\hat{x}_{A},\ \ \hat{p}_{B}^{\left(
0\right)  }e^{r}. \label{eq:qnd-no-loss}%
\end{equation}
She sends the second mode over a quantum channel to a receiver Bob.

The above operations satisfy the requirements for an $\left(  e^{-2r}\right)
$-approximate position-quadrature coherent channel. They satisfy constraint
(\ref{eq:position-retaining}) because the position quadrature of Alice's final
mode is equal to the position quadrature of Alice's original mode. They
satisfy constraint (\ref{eq:back-action}) because the momentum quadrature of
Alice's final mode is equal to the momentum quadrature of Alice's original
mode plus a noise term $-\hat{p}_{B}^{\left(  0\right)  }e^{r}$ so that
$\hat{p}_{\Delta_{X}}=-\hat{p}_{B}^{\left(  0\right)  }e^{r}$. They satisfy
constraint (\ref{eq:position-copying}) because the position quadrature of
Bob's final mode is equal to the position quadrature of Alice's original mode
plus a noise term $\hat{x}_{B}^{\left(  0\right)  }e^{-r}$ so that $\hat
{x}_{\Delta_{X}}=\hat{x}_{B}^{\left(  0\right)  }e^{-r}$.

Examination of the transformation in (\ref{eq:qnd-def}) confirms that the
above protocol also implements a nonlocal quantum nondemolition interaction.
This interpretation is rather obvious given that Alice performs the
interaction locally and sends one mode over a quantum channel.

Suppose we implement the needed quantum nondemolition interaction with FMA's
method. Then the coherent channel is $\epsilon$-approximate where%
\begin{equation}
\epsilon=\beta\left(  \left(  1-\eta\right)  /\eta T+e^{-2r}\right)  +e^{-2r}.
\end{equation}
The channel is useful if photodetector efficiency $\eta\rightarrow1$ and
squeezing strength $r$\ becomes large so that $\epsilon$ becomes small.

This particular method of implementing a coherent channel is somewhat wasteful
in its usage of resources because a quantum channel implements the coherent
channel. We later discuss two methods that make more efficient usage of
resources. The first method is coherent communication assisted by entanglement
and classical communication (CCAECC). The second is coherent superdense
coding. CCAECC uses one classical channel and one bipartite entangled state to
implement a coherent channel. Coherent superdense coding uses one bipartite
entangled state and one quantum channel to implement two coherent channels.

\subsection{Coherent Communication Assisted by Entanglement and Classical
Communication (CCAECC)}

Another method for implementing a coherent channel is CCAECC. It uses
bipartite entanglement and classical communication to implement a coherent
channel. We first give a brief description of the discrete-variable protocol
and follow with the continuous-variable description.

Suppose Alice has a qubit $\left\vert \psi\right\rangle ^{A}=\alpha\left\vert
0\right\rangle ^{A}+\beta\left\vert 1\right\rangle ^{A}$ that she wants to
send through a coherent channel. Suppose further that Alice and Bob possess an
ebit where Alice has qubit $A_{1}$ and Bob qubit $B$. Alice appends an ancilla
$\left\vert 0\right\rangle ^{A_{2}}$ and performs a CNOT\ between $A_{1}$ and
$A_{2}$. The resulting state is entangled so that the global state is as
follows:%
\begin{equation}
\left\vert \psi\right\rangle ^{A}\left(  \left\vert 0\right\rangle ^{A_{1}%
}\left\vert 0\right\rangle ^{A_{2}}\left\vert 0\right\rangle ^{B}+\left\vert
1\right\rangle ^{A_{1}}\left\vert 1\right\rangle ^{A_{2}}\left\vert
1\right\rangle ^{B}\right)  /\sqrt{2}.
\end{equation}
Alice performs a Bell measurement on qubits $A$ and $A_{1}$. She performs
teleportation-like corrective operations \cite{PhysRevLett.70.1895}\ on her
qubit $A_{2}$ using the two bits resulting from the Bell measurement. The
resulting state is a uniform mixture of the following two pure states:%
\begin{align}
&  \alpha\left\vert 0\right\rangle ^{A_{2}}\left\vert 0\right\rangle
^{B}+\beta\left\vert 1\right\rangle ^{A_{2}}\left\vert 1\right\rangle ^{B},\\
&  \alpha\left\vert 0\right\rangle ^{A_{2}}\left\vert 1\right\rangle
^{B}+\beta\left\vert 1\right\rangle ^{A_{2}}\left\vert 0\right\rangle ^{B}.
\end{align}
Alice sends Bob one bit so that he can perform a corrective Pauli $X$
operation. The resulting state is as follows%
\begin{equation}
\alpha\left\vert 0\right\rangle ^{A_{2}}\left\vert 0\right\rangle ^{B}%
+\beta\left\vert 1\right\rangle ^{A_{2}}\left\vert 1\right\rangle ^{B},
\end{equation}
so that Alice and Bob simulate a coherent channel from qubit $A_{1}$ to qubits
$A_{2}$ and $B$.

The continuous-variable method for CCAECC is similar to continuous-variable
teleportation \cite{prl1998braunstein}, a continuous-variable teleportation
network \cite{PhysRevLett.84.3482,nature2004furusawa}, and continuous-variable
quantum telecloning \cite{PhysRevLett.87.247901} though the protocol differs
from all of the above protocols in different ways. It differs from
continuous-variable teleportation because we use three-mode entanglement
shared among two parties. It differs from the teleportation network because we
have only two parties instead of three parties. It also differs from the
teleportation network because it implements a coherent channel rather than a
quantum channel. The protocol is perhaps most similar to quantum telecloning.
But it differs from it because we have two parties instead of three. The
entanglement in our protocol is GHZ-like, but quantum telecloning employs
W-like entanglement.%
\begin{figure}
[ptb]
\begin{center}
\includegraphics[
natheight=11.583300in,
natwidth=20.805700in,
height=2.1421in,
width=3.3313in
]%
{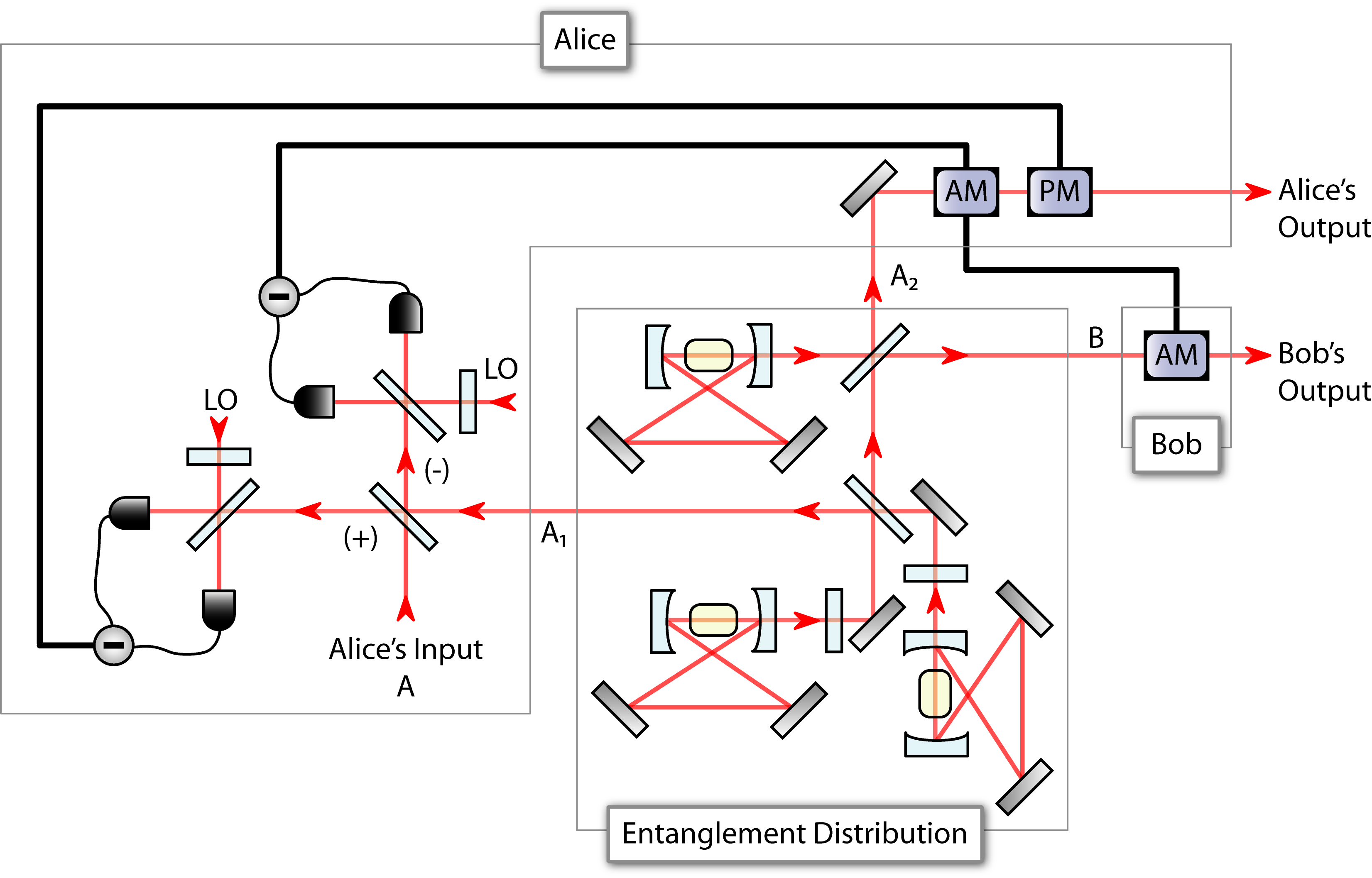}%
\caption{(Color online)\ Coherent channel implemented with entanglement and
classical communication. The protocol has similarities to previous protocols
\cite{prl1998braunstein,PhysRevLett.84.3482,nature2004furusawa,PhysRevLett.87.247901}%
, but has significant differences as well. Alice and Bob share a three-mode
entangled state. Alice possesses the first two modes and Bob possesses the
third. Alice has a mode $A$\ that she wants to send through a coherent
channel. She performs teleportation-like measurements on mode $A$\ and her
first mode in the entangled state. She performs feedforward control on her
second mode of the entangled state. She sends one variable over a classical
communications channel so that Bob can perform feedforward control. The
resulting state is equivalent to one sent through a coherent channel.}%
\label{fig:conat-GHZ}%
\end{center}
\end{figure}

Now let us describe the continuous-variable protocol. Suppose that Alice
possesses a mode $A$ that she wants to transmit through a coherent channel.
Alice and Bob possess a three-mode entangled state. Alice possesses the first
two modes $A_{1}$ and $A_{2}$ and Bob possesses the last mode $B$. We assume
that a three-port device called a \textquotedblleft tritter\textquotedblright%
\ creates the entanglement necessary for the protocol
\cite{PhysRevLett.84.3482,nature2004furusawa}. The entanglement that results
from the tritter is equivalent to the state from \cite{PhysRevLett.84.3482}.
The following correlations hold for modes $A_{1}$, $A_{2}$, and $B$%
\begin{align}
\hat{x}_{A_{1}}-\hat{x}_{A_{2}}  &  =\sqrt{3/2}e^{-r}\hat{x}_{2}^{\left(
0\right)  }-\sqrt{2^{-1}}e^{-r}\hat{x}_{3}^{\left(  0\right)  }\nonumber\\
\hat{x}_{A_{2}}-\hat{x}_{B}  &  =\sqrt{2}e^{-r}\hat{x}_{3}^{\left(  0\right)
}\nonumber\\
\hat{p}_{A_{1}}+\hat{p}_{A_{2}}+\hat{p}_{B}  &  =\sqrt{3}e^{-r}\hat{p}%
_{1}^{\left(  0\right)  } \label{eq:GHZ1-ent}%
\end{align}
where $\hat{x}_{2}^{\left(  0\right)  }$ and $\hat{x}_{3}^{\left(  0\right)
}$ are the position quadratures of the second and third modes sent through the
tritter and $\hat{p}_{1}^{\left(  0\right)  }$ is the momentum quadrature of
the first mode sent through the tritter. Parameter $r$ denotes the squeezing
strength of the original three vacuum modes sent through the tritter. We
assume that the squeezing strength is the same for all three vacuum modes.

Figure~\ref{fig:conat-GHZ}\ outlines the optical circuit needed for our
protocol. It begins by Alice mixing her modes $A$ and $A_{1}$\ at a
beamsplitter. The output Heisenberg-picture observables $\hat{x}_{\pm}$,
$\hat{p}_{\pm}$ are as follows:%
\begin{align*}
\hat{x}_{\pm} &  =\left(  \hat{x}_{A}\pm\hat{x}_{A_{1}}\right)  /\sqrt{2},\\
\hat{p}_{\pm} &  =\left(  \hat{p}_{A}\pm\hat{p}_{A_{1}}\right)  /\sqrt{2}.
\end{align*}
The observables of modes $A_{2}$ and $B$ are then as follows:%
\begin{align*}
\hat{x}_{A_{2}} &  =\hat{x}_{A}-\left(  \hat{x}_{A_{1}}-\hat{x}_{A_{2}%
}\right)  -\sqrt{2}\hat{x}_{-},\\
\hat{x}_{B} &  =\hat{x}_{A}-\left(  \hat{x}_{A_{1}}-\hat{x}_{B}\right)
-\sqrt{2}\hat{x}_{-},\\
\hat{p}_{A_{2}} &  =\hat{p}_{A}+\left(  \hat{p}_{A_{1}}+\hat{p}_{A_{2}}%
+\hat{p}_{B}\right)  -\hat{p}_{B}-\sqrt{2}\hat{p}_{+}.
\end{align*}
Alice performs a position-quadrature homodyne detection on mode ($-$) and a
momentum-quadrature homodyne detection on mode ($+$). Suppose the
photodetectors have efficiency $\eta$. The observables $\hat{x}_{-}$ and
$\hat{p}_{+}$ collapse to values $x_{-}$ and $p_{+}$ respectively. Alice
modulates mode $A_{2}$ locally by displacing the position quadrature by
$\sqrt{2}x_{-}$ and the momentum quadrature by $\sqrt{2}p_{+}$. She sends the
value $x_{-}$ over a classical communications channel. Bob displaces his
position quadrature by an amount $\sqrt{2}x_{-}$. Let us call the resulting
modes $A^{\prime}$ and $B^{\prime}$. The Heisenberg-picture observables are
then as follows after they perform the above operations:%
\begin{align*}
\hat{x}_{A^{\prime}} &  =\hat{x}_{A}-\left(  \hat{x}_{A_{1}}-\hat{x}_{A_{2}%
}\right)  -\sqrt{2\left(  1-\eta\right)  /\eta}\hat{x}_{1}^{\left(  0\right)
},\\
\hat{p}_{A^{\prime}} &  =\hat{p}_{A}+\left(  \hat{p}_{A_{1}}+\hat{p}_{A_{2}%
}+\hat{p}_{B}\right)  -\hat{p}_{B}+\sqrt{2\left(  1-\eta\right)  /\eta}\hat
{p}_{2}^{\left(  0\right)  },\\
\hat{x}_{B^{\prime}} &  =\hat{x}_{A}-\left(  \hat{x}_{A_{1}}-\hat{x}%
_{B}\right)  -\sqrt{2\left(  1-\eta\right)  /\eta}\hat{x}_{1}^{\left(
0\right)  },\\
\hat{p}_{B^{\prime}} &  =\hat{p}_{B}.
\end{align*}
The quadrature operators $\hat{x}_{1}^{\left(  0\right)  }$ and $\hat{p}%
_{2}^{\left(  0\right)  }$ are independent and thus commuting observables.
They have the fluctuations of the vacuum and model the inefficiency that both
homodyne detectors introduce. Let us determine if the above operations
implement a coherent channel. Squeezing introduces some extra noise in mode
$A_{2}$. The difference between the position quadratures of modes $A_{2}$ and
$B$ is as follows:%
\[
\left\langle \left(  \hat{x}_{A^{\prime}}-\hat{x}_{B^{\prime}}\right)
^{2}\right\rangle =2e^{-2r}.
\]
We subtract the original momentum of Alice's mode $A$ from the total momentum
of modes $A^{\prime}$ and $B^{\prime}$ when considering coherent channel
performance according to the performance measure in
(\ref{eq:coherent-performance}). The quantity $\hat{p}_{A^{\prime}}+\hat
{p}_{B^{\prime}}-\hat{p}_{A}$ is as follows:%
\[
\left\langle \left(  \hat{p}_{A^{\prime}}+\hat{p}_{B^{\prime}}-\hat{p}%
_{A}\right)  ^{2}\right\rangle =3e^{-2r}+2\left(  1-\eta\right)  /\eta.
\]
The above operations therefore implement a $(3e^{-2r}+2\left(  1-\eta\right)
/\eta)$-approximate position-quadrature coherent channel. The channel becomes
ideal as squeezing strength $r$ becomes large and as photodetector efficiency
approaches unity.

We can also view the above operations as performing a nonlocal quantum
nondemolition interaction. The information in the original position quadrature
$\hat{x}_{A}$ transfers to both modes $A^{\prime}$ and $B^{\prime}$ with the
addition of some noise. Filip's scheme requires bipartite entanglement and
two-way classical communication \cite{filip:052313}. Our method requires
bipartite entanglement and one-way classical communication. Our scheme is thus
an improvement if we view communication as expensive and local operations as free.

\subsection{Coherent Superdense Coding}%

\begin{figure*}
\begin{center}
\includegraphics[
natheight=10.430600in,
natwidth=27.561300in,
height=2.5712in,
width=6.7896in
]
{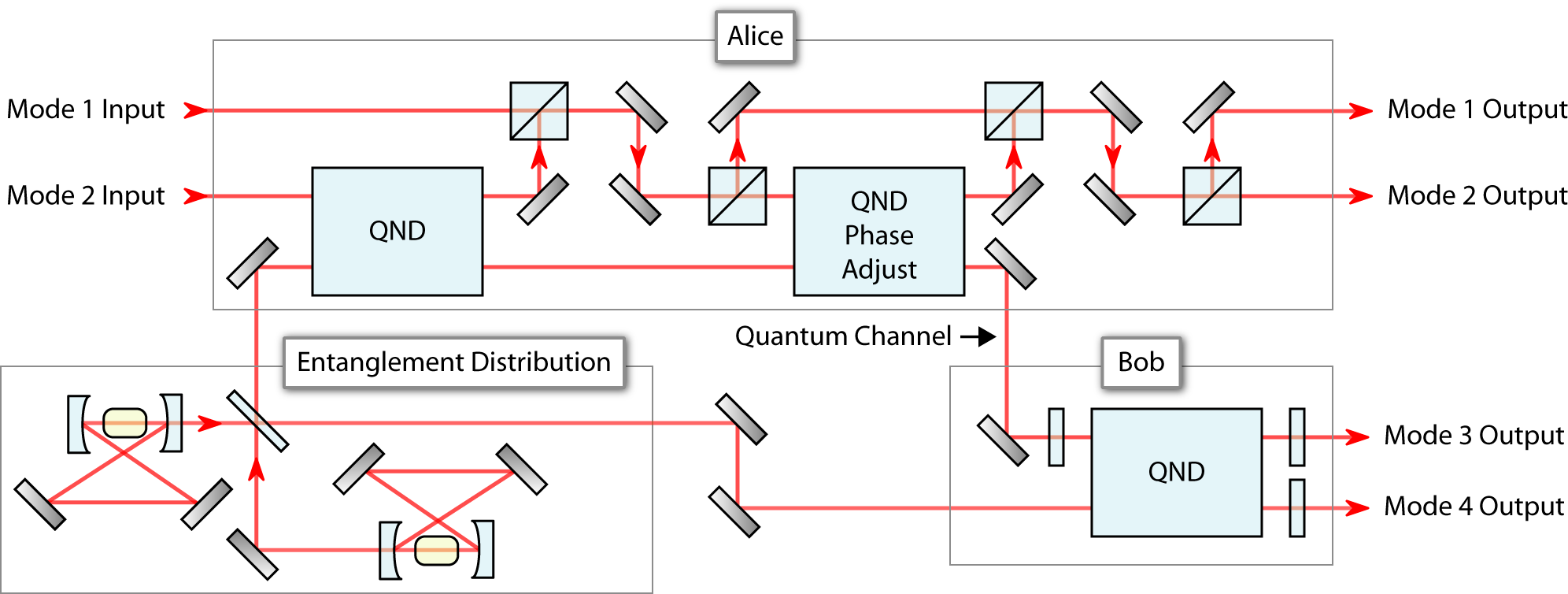}
\caption
{(Color online) The above linear-optical circuit implements coherent superdense coding
or, equivalently, two-nonlocal quantum nondemolition interactions.
All phase shifters in the above circuit rotate by $\pi$.
Alice possesses two
modes that she wants to send through a coherent channel. Alice and Bob
share two entangled modes. Alice performs some local operations on her two modes
and her half of the entangled state. She sends one mode over a quantum channel to Bob.
Bob performs some local operations. The resulting states are equivalent to those that
would result from Alice sending her two modes through two coherent channels. The states
are also equivalent to those that would result from performing two nonlocal quantum
nondemolition interactions.}
\label{fig:coherent-dense}
\end{center}
\end{figure*}%
Braunstein and Kimble proposed a theoretical method for performing
continuous-variable superdense coding with linear optics
\cite{pra2000braunstein}. Later work outlined how to make their protocol
coherent \cite{wilde:060303}. We review the theoretical operation of coherent
superdense coding briefly and follow with an analysis of losses incurred by
employing Filip et al's technique for a quantum nondemolition interaction.
Figure~\ref{fig:coherent-dense} gives a way to implement coherent superdense
coding experimentally. We also provide two observations concerning coherent
superdense coding.

Suppose Alice possesses two modes $1$ and 2 at the beginning of the protocol.
Alice and Bob also share a two-mode entangled state with Alice having mode
three and Bob mode four. The second moment noise of the quadrature observables
obey the following inequalities:%
\[
\langle\left(  \hat{x}_{3}-\hat{x}_{4}\right)  ^{2}\rangle,\ \ \ \langle
\left(  \hat{p}_{3}+\hat{p}_{4}\right)  ^{2}\rangle\leq\delta.
\]
Modes three and four are entangled if $\delta<1$ \cite{prl2000lmduan}. Alice
couples her second mode and her half of the entangled state in a quantum
nondemolition interaction. She swaps her first mode with her second mode. She
couples the second mode with the third mode in a quantum nondemolition
interaction. She adjusts the phases of the local oscillators so that the
momentum quadrature copies rather than the position quadrature. She sends her
third mode over a quantum channel to Bob and swaps her first mode with her
second mode. Bob couples the received mode and his half of the entangled state
in a quantum nondemolition interaction. This quantum nondemolition interaction
subtracts the position quadrature rather than adding it because of the $\pi$
phase shifters acting on his received mode. He finally sends his last mode
through a $\pi$ phase shifter to reflect its position and momentum quadrature.
The resulting four modes have the following Heisenberg-picture observables:%
\begin{align}
\hat{x}_{1}^{\prime} &  =\hat{x}_{1}-\left(  \hat{x}_{2}+\hat{x}_{3}\right)
,\ \ \hat{p}_{1}^{\prime}=\hat{p}_{1},\nonumber\\
\hat{x}_{2}^{\prime} &  =\hat{x}_{2},\ \ \hat{p}_{2}^{\prime}=\hat{p}_{2}%
-\hat{p}_{3},\nonumber\\
\hat{x}_{3}^{\prime} &  =\hat{x}_{2}+\hat{x}_{3},\ \ \hat{p}_{3}^{\prime}%
=\hat{p}_{1}+\left(  \hat{p}_{3}+\hat{p}_{4}\right)  ,\nonumber\\
\hat{x}_{4}^{\prime} &  =\hat{x}_{2}+\left(  \hat{x}_{3}-\hat{x}_{4}\right)
,\ \ \hat{p}_{4}^{\prime}=-\hat{p}_{4}.
\end{align}
Modes one and three satisfy the conditions for an $\delta$-approximate
momentum-quadrature coherent channel. Modes two and four satisfy the
conditions for an $\delta$-approximate position-quadrature coherent channel.
Let us stress that the above operations assume ideal quantum nondemolition interactions.

Consider the effect of using FMA's scheme to implement each quantum
nondemolition interaction. Each quantum nondemolition interaction adds noise
from two squeezed sources and two vacuum sources due to inefficient homodyne
detection. Parameter $\eta_{F}$ in (\ref{eq:QND-noise-bound}) bounds the noise
that each quantum nondemolition interaction adds. We use three quantum
nondemolition interactions in the above protocol. All three quantum
nondemolition interactions affect the observables present in the final output
modes three and four. Their effects are independent and additive. Therefore
using FMA's quantum nondemolition interaction gives two $\left(  \delta
+3\eta_{F}\right)  $-approximate coherent channels. These coherent channels
are useful if $\delta+3\eta_{F}$ is small---less than 1/2 or 1 depending on
which teleportation bound we want to surpass.

We make several observations about the above protocol. It reduces to ordinary
(incoherent) superdense coding \cite{pra2000braunstein} when Alice uses
certain input states. Suppose she encodes two classical variables $p$ and $x$
in modes one and two respectively. She performs this encoding by making modes
one and two be highly squeezed in the momentum quadrature and the position
quadrature respectively. Modes one and two approach a momentum-quadrature
eigenstate $\left\vert p\right\rangle $ and a position-quadrature eigenstate
$\left\vert x\right\rangle $ in the infinite squeezing limit. Then Bob's modes
three and four are approximately a momentum-quadrature $\left\vert
p\right\rangle $ and a position-quadrature eigenstate $\left\vert
x\right\rangle $. The noise from finite-squeezing and the quantum
nondemolition interactions affect his states so that they are not perfect
eigenstates. Bob can perform a measurement to retrieve an approximation of the
two classical variables $p$ and $x$ that Alice first sent. Thus this protocol
reduces to incoherent dense coding in this sense.

Another way of viewing this protocol is that it implements two nonlocal
quantum nondemolition\ interactions with three local quantum
nondemolition\ interactions. We can view the results as quantum nondemolition
interactions because we transfer information about the momentum quadrature
$\hat{p}_{1}$ to mode three and we transfer information about the position
quadrature $\hat{x}_{2}$ to mode four. The above coherent superdense coding
protocol uses one quantum channel and one set of entangled modes to achieve
two nonlocal quantum nondemolition interactions. Coherent superdense coding
then is an interesting way to implement two nonlocal quantum nondemolition
interactions if we view communication and entanglement as expensive and local
operations as free.

\section{Coherent Teleportation}

\label{sec:coh-tele}
\begin{figure*}
\begin{center}
\includegraphics[
natheight=9.1900in,
natwidth=30.6200in,
height=2.0702in,
width=6.2029in
]
{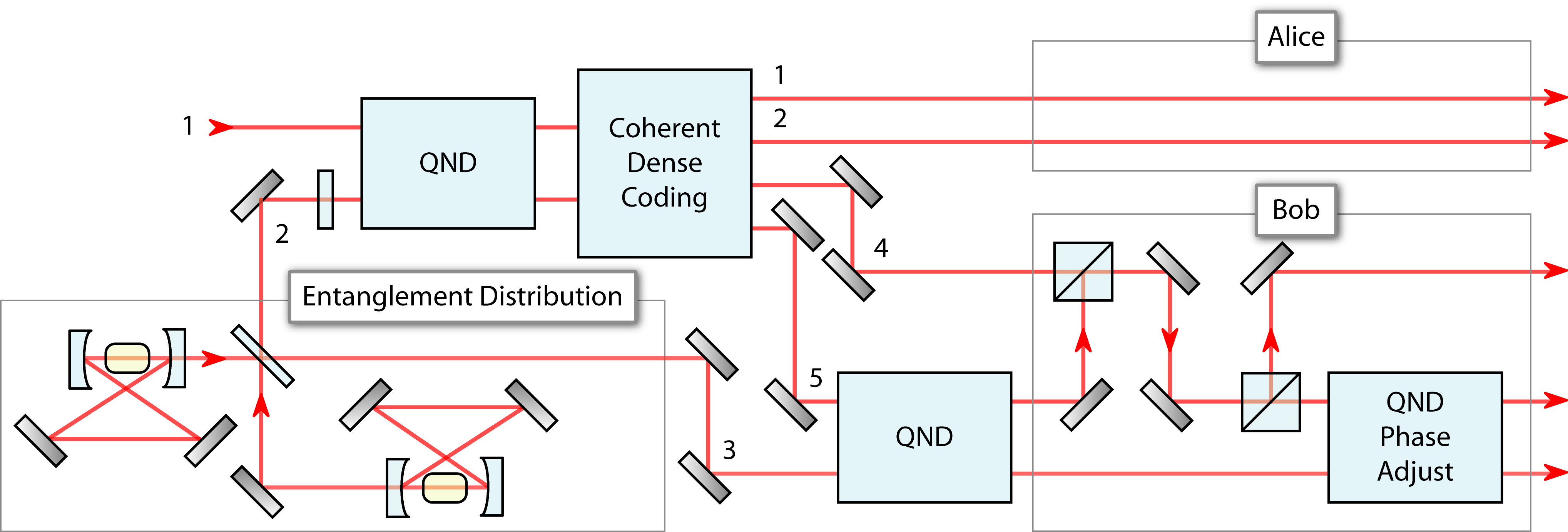}
\caption
{(Color online) The above circuit implements the coherent teleportation protocol with a linear-optical circuit.
All phase shifters in the above circuit rotate by $\pi$.
Alice possesses one mode that she wants to teleport and shares an entangled state with Bob. Alice
performs some local operations and sends her two modes through the circuit for coherent superdense coding.
Coherent superdense coding gives Alice two modes and Bob two modes. Bob performs some local operations.
The result is that Alice's original mode teleports to mode three. Alice and Bob additionally share two
entangled states at the end of the protocol.}
\label{fig:coherent-tele}
\end{center}
\end{figure*}

Braunstein and Kimble also gave a theoretical proposal for
performing linear-optical continuous-variable teleportation
\cite{pra2000braunstein}. The later work in \cite{wilde:060303}\ illustrated
how to make their protocol coherent. We review the operation of coherent
teleportation and give a loss analysis when employing FMA's scheme. We also
suggest an experimental method with Bell inequalities to determine if coherent
teleportation is successful. Figure~\ref{fig:coherent-tele} gives a way to
implement coherent teleportation experimentally with linear optics.

Coherent teleportation is a two-party protocol. Alice possesses a mode
one\ that she wants to teleport to Bob. She shares two continuous-variable
entangled states with Bob. We label the modes in the first pair as two and
three. Alice possesses mode two and Bob possesses mode three. The outputs of
the offline squeezers in Figure~\ref{fig:coherent-tele} couple at a
beamsplitter to give the first pair of entangled modes. We label the modes in
the second pair as four and five. The coherent dense coding circuit requires
an entangled pair so modes four and five serve this purpose. Alice possesses
mode four and Bob possesses mode five. Both entangled pairs have the following
correlations:%
\begin{align*}
\langle\left(  \hat{x}_{2}-\hat{x}_{3}\right)  ^{2}\rangle,\ \ \langle\left(
\hat{p}_{2}+\hat{p}_{3}\right)  ^{2}\rangle &  \leq\delta,\\
\langle\left(  \hat{x}_{4}-\hat{x}_{5}\right)  ^{2}\rangle,\ \ \langle\left(
\hat{p}_{4}+\hat{p}_{5}\right)  ^{2}\rangle &  \leq\delta.
\end{align*}
The above states are entangled if $\delta<1$.%

The protocol begins with Alice sending mode two through a $\pi$ phase shifter.
She couples her modes one and two in a quantum nondemolition interaction. She
sends her two modes through the coherent superdense coding circuit in
Figure~\ref{fig:coherent-dense}. Recall that the first two outputs of the
coherent superdense coding circuit belong to Alice and the second two to Bob.
Bob combines his mode five from coherent superdense coding and mode three in a
quantum nondemolition interaction. He swaps his mode four with the first
output of the quantum nondemolition interaction. He then couples his mode four
and three in a quantum nondemolition interaction. He adjusts the phases of the
local oscillators so that it copies the momentum quadrature rather than the
position quadrature. The relations between the input Heisenberg-picture
observables and the output observables are as follows:%
\begin{align*}
\hat{x}_{1}^{\prime} &  =\hat{x}_{2}-\hat{x}_{4},\ \ \hat{p}_{1}^{\prime}%
=\hat{p}_{2}+\hat{p}_{1},\\
\hat{x}_{2}^{\prime} &  =\hat{x}_{1}-\hat{x}_{2},\ \ \hat{p}_{2}^{\prime
}=-\hat{p}_{2}-\hat{p}_{4},\\
\hat{x}_{3}^{\prime} &  =\hat{x}_{1}+\left(  \hat{x}_{3}-\hat{x}_{2}\right)
+\left(  \hat{x}_{4}-\hat{x}_{5}\right)  ,\\
\hat{p}_{3}^{\prime} &  =\hat{p}_{1}+\left(  \hat{p}_{2}+\hat{p}_{3}\right)
+\left(  \hat{p}_{4}+\hat{p}_{5}\right)  ,\\
\hat{x}_{4}^{\prime} &  =\hat{x}_{5}-\hat{x}_{3},\ \ \hat{p}_{4}^{\prime}%
=\hat{p}_{2}+\hat{p}_{1}+\left(  \hat{p}_{4}+\hat{p}_{5}\right)  ,\\
\hat{x}_{5}^{\prime} &  =\hat{x}_{1}-\hat{x}_{2}+\left(  \hat{x}_{4}-\hat
{x}_{5}\right)  ,\ \ \hat{p}_{5}^{\prime}=-\hat{p}_{5}-\hat{p}_{3}.
\end{align*}
The state in mode one teleports to mode three. The average fildelity $F$\ for
teleporting a coherent state \cite{science1998furusawa} is as follows%
\begin{equation}
F=2/\left[  \left(  \left\langle \left(  \Delta\hat{x}_{\text{tel}}\right)
^{2}\right\rangle +1\right)  \left(  \left\langle \left(  \Delta\hat
{p}_{\text{tel}}\right)  ^{2}\right\rangle +1\right)  \right]  ^{1/2},
\end{equation}
where $\hat{x}_{\text{tel}}$ and $\hat{p}_{\text{tel}}$ are the quadratures of
the teleported mode. The fidelity $F$ for coherent teleportation is thus%
\begin{equation}
F=1/\left(  1+\delta\right)  .
\end{equation}
Modes one and four are entangled and modes two and five are also entangled
because the following correlations hold:%
\begin{align}
\left\langle \left(  \hat{x}_{1}^{\prime}+\hat{x}_{4}^{\prime}\right)
^{2}\right\rangle ,\ \ \left\langle \left(  \hat{p}_{1}^{\prime}-\hat{p}%
_{4}^{\prime}\right)  ^{2}\right\rangle  &  \leq\delta,\label{eq:entang-corr}%
\\
\left\langle \left(  \hat{x}_{2}^{\prime}-\hat{x}_{5}^{\prime}\right)
^{2}\right\rangle ,\ \ \left\langle \left(  \hat{p}_{2}^{\prime}+\hat{p}%
_{5}^{\prime}\right)  ^{2}\right\rangle  &  \leq\delta.\nonumber
\end{align}

Consider if we employ FMA's scheme to implement the quantum nondemolition
interactions in coherent teleportation. Recall that this scheme adds at most
$3\eta_{F}$ to the second moments of the quadratures in coherent superdense
coding. Coherent teleportation requires three extra quantum nondemolition
interactions. These interactions each add at most a noise factor of $\eta_{F}$
to the second moments of Bob's output quadrature variables. Thus the total
noise contribution from six quantum nondemolition interactions is no more than
$6\eta_{F}$. This contribution affects the fidelity of teleportation by
bounding it as follows:%
\begin{equation}
F>1/\left(  1+\delta+6\eta_{F}\right)  .
\end{equation}
The noise also affects the entanglement correlations. The quantity
$\delta+6\eta_{F}$ upper bounds each correlation term in (\ref{eq:entang-corr}%
). FMA's scheme then implements coherent teleportation if the quantity
$6\eta_{F}$ is small enough so that either $F>1/2$ or $F>2/3$ and each
entanglement correlation in (\ref{eq:entang-corr}) is less than one.

Coherent teleportation preserves two sets of entanglement. An experimentalist
implementing coherent teleportation should verify that the resulting output
entanglement violates a Bell inequality. One suitable test uses photon number
parity measurements for a Bell test \cite{PhysRevA.58.4345}. This test should
determine if the entanglement is useful for quantum communication.

Let us examine coherent teleportation in more detail. Two entangled modes and
a quantum channel implement the same resources. Why not just use the quantum
channel to \textquotedblleft teleport\textquotedblright\ the first mode and
leave the other entangled modes as they are?\ Let us stress that one purpose
of our coherent teleportation protocol is to illustrate the capabilities of a
coherent channel. Coherent teleportation is one example of a useful protocol
that we can implement given that two coherent channels are available as a
resource. The implementation of these coherent channels may be achieved with
any of the methods outlined in this paper or with an alternate method of which
we are not yet aware.

\section{Conclusion}

Each protocol in this work has an implementation with linear optics. FMA's
scheme for a quantum nondemolition interaction aids in realizing the
linear-optical circuits. Our analysis of noise accumulation should prove
useful for determining the realistic performance of each protocol.

We give a different interpretation to a coherent channel in this work. It is a
special case of a nonlocal quantum nondemolition interaction that has unity
gain and the correlations in\ (\ref{eq:position-copying}%
--\ref{eq:coherent-performance}). We present three different methods for
implementing a coherent channel. These methods are equivalent to nonlocal
quantum nondemolition interactions. Two of these methods---CCAECC and coherent
superdense coding---offer an improvement over previous methods for a nonlocal
interaction if we view communication as expensive and local operations as free.

\section{Acknowledgements}

The authors thank Igor Devetak, Radim Filip, Aram Harrow, Hari Krovi, and Jon Yard for useful comments.
MMW acknowledges support from NSF Grants 0545845 and CCF-0448658. TAB
acknowledges support from NSF Grant CCF-0448658. All authors acknowledge
support from the Army Research Office and the Disruptive Technology Office.

\bibliographystyle{apsrev}
\bibliography{conat-exp}

\end{document}